\documentclass{aa}

\usepackage{amsmath}
\usepackage{siunitx}
\usepackage{xfrac}
\usepackage{natbib}
\usepackage{xcolor}
\usepackage{graphicx}
\usepackage{booktabs}
\usepackage{csquotes}
\usepackage{subcaption}
\usepackage{multirow,array}
\makeatletter
  \renewcommand*\aa@pageof{, page \thepage{} of \pageref*{LastPage}}
\makeatother
\usepackage{hyperref}            

\hypersetup{pdfpagemode = {UseNone},
            pdftitle = {1to1},
            pdfcreator = {\LaTeX},
            pdfproducer = {pdfeTeX-0.\the\pdftexversion\pdftexrevision},
            pdfauthor = {Stefan Adelbert, Anna Penzlin},
            pdfsubject = {},
            pdfview = {FitH},
            pdfstartview = {FitH},
            colorlinks = {true},
            linkcolor = [rgb]{0,0.35,0.7},
            citecolor = [rgb]{0,0.35,0.7},
            filecolor = [rgb]{0.61,0,0},
            urlcolor = [rgb]{0.61,0,0},
           }

\usepackage[normalem]{ulem}
           
\usepackage{bm}

\colorlet{darkgreen}{green!60!black}

\colorlet{darkpink}{pink!70!black}

\newcommand{\beq}{\begin{equation}}
\newcommand{\eeq}{\end{equation}}

\authorrunning{Adelbert et al.}

\begin{document}

\title{Stability of coorbital planets around binaries}

\author{Stefan Adelbert\inst{\ref{inst1}}, 
		Anna~B.~T.~Penzlin\inst{\ref{inst1},\ref{inst2}},
        Christoph~M.~Sch\"afer\inst{\ref{inst1}},
        Wilhelm Kley\thanks{passed away}\inst{\ref{inst1}},
        Billy Quarles\inst{\ref{inst3}},
        Rafael Sfair\inst{\ref{inst1},\ref{inst4}}
        }

\institute{
Institut f\"ur Astronomie und Astrophysik, Universität T\"ubingen,
Auf der Morgenstelle 10, 72076 T\"ubingen, Germany\label{inst1}
\and
Astrophysics Group, Department of Physics, Imperial College London, Prince Consort Rd, London, SW7 2AZ, UK \label{inst2}
\and
Department of Applied Mathematics and Physics, Valdosta State University, Valdosta GA, 31698, USA \label{inst3}
\and
Grupo de Dinâmica Orbital e Planetologia, São Paulo State University, UNESP, Guaratinguetá, CEP 12516-410, São Paulo, Brazil \label{inst4}\\
\email {a.penzlin@imperial.ac.uk, stefan.adelbert@uni-tuebingen.de}\\
}

\date{}

\abstract
{
In previous hydrodynamical simulations, we found a mechanism for nearly circular binary stars, like Kepler-413, to trap two planets in a stable 1:1 resonance. Therefore, the stability of coorbital configurations becomes a relevant question for planet formation around binary stars. Here, we investigate the coorbital planet stability using a Kepler-413 analogue as an example and then expand the parameters to  study general n-body stability of planet pairs in eccentric horseshoe orbits around binaries. The stability is tested by evolving the planet orbits for $10^5$ binary periods with varying initial semi-major axes and planet eccentricities. 
The unstable region of a single circumbinary planet is used as a comparison to the investigated coorbital configurations in this work. We confirm previous findings on the stability of single planets and find a first order linear relation between orbit eccentricity $e_\mathrm{p}$ and pericentre to identify stable orbits for various binary configurations. Such a linear relation is also found for the stability of 1:1 resonant planets around binaries. Stable orbits for eccentric horseshoe configurations exist with a pericentre closer than seven binary separations and, in the case of Kepler-413, the pericentre of the first stable orbit can be approximated by $r_\mathrm{c,peri}= (2.90\, e_\mathrm{p} + 2.46)\, a_\mathrm{bin}$.
}

\keywords{
          N-body --
          Binaries: general --
          Planets and satellites: resonance --
         }

\maketitle

\section{Introduction}\label{sec:intro}
Coorbital configurations or 1:1 resonances could thus far only be detected in small objects in the Solar System (e.g.,\ Janus and Epimetheus) possibly also due to detection biases \citep{2012Giuppone}.  
Theoretically, around a single star \cite{2018CeMDA.130...24L} found stable 1:1 resonances for planets with masses of $10^{-5}\,M_\star$ for orbit eccentricities up to 0.5. \cite{2023Raymond} recently demonstrated that even multi-planet coorbital configurations can be stable for low masses. \cite{2006A&A...450..833C} found in single-star systems that the migration of protoplanets can lead to 1:1 resonant horseshoe orbits. However, to achieve stability in the single star case the resonance with a larger additional planet is needed \citep{2009CresswellNelson, 2019arXiv190107640L}. Without the additional resonant planet continued migration leads to increasing libration and breaking of the resonance. 
Recently, \cite{2023L5PDS70} found evidence for accumulated material at the Lagrange L5 point of the orbit of the embedded planet PDS 70 b. This could indicate the formation of coorbital Trojan objects in protoplanetary systems as theoretically predicted.

In binary star systems, migration of the planet through the circumbinary disc can be halted by the eccentric inner cavity in the disc caused by the binary \citep{2017Mutter,2019Kley,2021Penzlin}. The eccentricity of the inner disc near the cavity refills planet-opened gaps and thereby hinders planets from reaching the inner cavity.
The inner circumbinary disc is most eccentric for low (<0.05) and high (>0.3) binary eccentricities \citep{2017Thun,2020Ragusa}.
This leads to the condition that allows a 1:1 resonance to form in a circumbinary disc of nearly circular binaries.
In \cite{2019Penzlin}, we found that a 1:1 resonance can form between comparable mass planets of $\approx10^{-5}$--$10^{-4}~M_\mathrm{bin}$ inside the circumbinary disc.
The migration of the planets is halted at comparable positions due to the structure of the inner disc and the damping forces of the eccentric disc allow them to enter a coorbital configuration.
The 1:1 resonance in the eccentric orbit, which resulted from the shape of the eccentric disc, remains stable even without the disc in n-body simulations. 
The n-body models, including a prescribed disc by \cite{2023Coleman}, also find instances of dynamically created coorbital planet configurations for Kepler-16 and Kepler-34 analogues \citep{2011Kepler16, 2012Kepler34-35}. 
During the embedded phase, the structure of the circumbinary disc leads to slow planet migration, exposing them to potentially unstable resonances with the binary on their orbit as studied in \cite{2022Martin} for single planets. 

In the case of a coorbital resonance, the destabilizing resonances may be even more important as the two planets librate around their common orbit and thereby enter mean motion resonant orbits with the binary even at larger distances from the resonant location \citep[see also][for resonant multi-planets]{2019Sutherland}.
Additionally, the resonant location widens with increased orbit eccentricity \citep{2006Mudryk}.
This prompts the question of which other circumbinary orbits are theoretically stable for 1:1 resonant planets. 

In the last decade, a dozen close circumbinary planets have been detected with the latest confirmed observation of TIC 172900988b \citep{2021TIC} and TOI-1338 b\&c \citep{2023TOI1338}.
Most detected single circumbinary planets show comparable orbits in units of binary distance close to the instability limit \citep{1986Dvorak,1999Holman}.
Studies by \cite{2016Popova, 2018Quarles, 2019Chen} have shown that also eccentric single planets close to the binary can be stable, and described how the closest stable orbit depends on the planet's eccentricity.

One special case is the Kepler-47 system that hosts three circumbinary planets, demonstrating that multi-planet configurations around binaries are possible \citep{2019Kepler47}.
In \cite{2019Penzlin} we found stable 1:1 resonant configurations for Kepler-47 and -413 analogues in hydrodynamic and n-body simulations. 
With this idea, \cite{2021Sudol} constrained the observability of resonant planets.

A theoretical study by \cite{1989JApA...10..347M} found only stable 1:1 resonances around binaries for zero-mass planets. 
However, since we already identified at least one stable orbit around a circular binary system with planet masses of $\leq 0.21\,M_\mathrm{jup}$, we investigate here stable planet orbits over the full range of planetary eccentricities with \texttt{rebound} n-body simulations \citep{2012Rein, 2015Rein} 
to identify the stability limit of 1:1-resonant, circumbinary planets with a Kepler-413 like host binary and for a range of binary parameters relevant to the known circumbinary systems.

Section \ref{sec:model} explains the n-body model used to analyse the orbit stability. In section \ref{sec:example}, Kepler 413 is used as an example to test the best set-up, explore the dynamics of the planet systems and establish a simple first order stability fitting function. We discuss the resulting parameter range of 1:1 resonant stable planet orbits for different binaries in section \ref{sec:resonance}. Finally, we discuss the results in \ref{sec:discuss} and summarize the findings in \ref{sec:results}.

\section{Models}\label{sec:model}
\begin{figure}
    \centering
    \resizebox{\hsize}{!}{\includegraphics{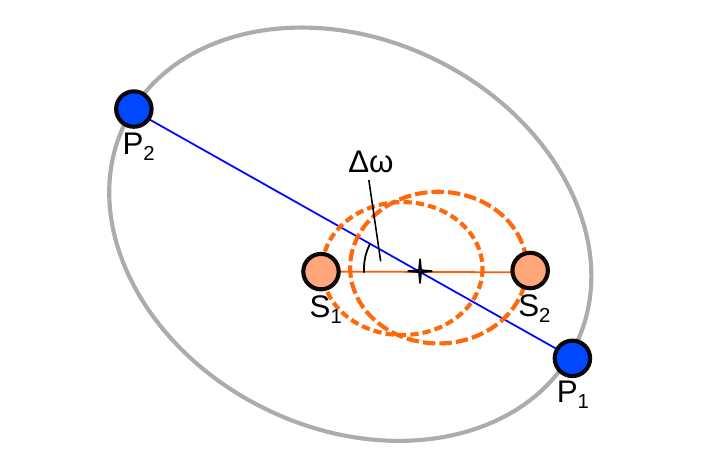}}
    \caption{Initial configuration of the stars ($S_1$ and $S_2$) and planets ($P_1$ and $P_2$) with the relative argument of periastron $\Delta\omega$ between planet and binary orbits.
    \label{img:sketch}
    }
\end{figure}

In this study, we investigate a binary system hosting circumbinary planets in an initial coorbital configuration (see Fig.~\ref{img:sketch}). We will compare these to systems containing only a single circumbinary planet. The initial orbit parameters of the n-body problem, the variations of parameters, and the simulation setup are defined as follows.

\subsection*{Binary parameters}

We will use the Kepler-413 binary as a fiducial case to go onto exploring a range of binary parameters.
We base the fiducial simulation on the parameters found by \citet{2014Kepler413}, therefore the binary mass ratio is $\mu_\mathrm{bin}=M_2/M_\mathrm{bin}=0.4$ using the primary mass $M_1$ and total binary mass $M_\mathrm{bin}$ and the binary eccentricity is $e_\mathrm{bin}=0.04$. Mass and distance are scalable with the binary parameters used, so we use the binary semi-major axis $1\,a_\mathrm{bin}$ for one standard length unit and $1\,M_\mathrm{bin}$ for the standard mass unit. The gravitational constant is normalized to 1, which results in a binary period of $T_\mathrm{bin}= 2\pi$, which we use as the standard time unit. The eccentricity of the planets varies between 0.0--0.9 in the stability analysis.
The setup is summarized in Table~\ref{tab:standard} as fiducial.

Investigating the binary parameters, the mass ratio $\mu_\mathrm{bin}$ takes the values [0.1, 0.2, 0.3, 0.4, 0.5] and the eccentricity $e_\mathrm{bin}$ [0.0, 0.125, 0.25, 0.375, 0.5] at the same time. This creates a $5 \times 5$ parameter space to understand the rudimentary effect of the binary on the planet's stable orbits(see "par" in Table~\ref{tab:standard}).
The binary orbit starts at its apocentre position.

\begin{table}[t]
\centering
\begin{tabular}{|c|c|c|c|c|c|c|}
\hline
 & $\mu_\mathrm{bin}$  &  $e_\mathrm{bin}$ & $ m_p$ &  $\Delta e_\mathrm{p}$  &$\Delta r_\mathrm{peri}$& $\Delta\omega$ \\
\hline
fid & 0.4    &  0.04  & $10^{-4}$ &  0.01  & $0.01$ & $\pi/2$  \\
par &0.1-0.5& 0.0-0.5& $10^{-4}$ &  0.01  & $0.01$ & $\pi/2$  \\
\hline
\end{tabular}
\caption{Summary of the physical parameters for the models. Binary mass ratio $\mu_\mathrm{bin}=M_2/M_\mathrm{bin}$ and eccentricity $e_\mathrm{bin}$ are chosen from Kepler-413 for the fiducial model "fid" and vary for the parameter investigation "par". 
The overall planet mass $m_\mathrm{p}$ is stated in units of $M_\mathrm{bin}$.
The planets are initialized with orbital eccentricities $e_\mathrm{p}$ ranging from 0.0-0.9 with a step width of $\Delta e_\mathrm{p}$ and a varying distance at pericentre with a step width of $\Delta r_\mathrm{peri}$ in units of $a_\mathrm{bin}$. The relative argument of periastron between planet and binary orbit $\Delta\omega$ was set to the least stable value.
}
\label{tab:standard}
\end{table}

\subsection*{Planet parameters}

The planetary orbits are initialized relative to the binary center of mass so that the relative argument of pericentre $\Delta \omega = \omega_{\rm bin} - \omega_p$ defines the relative alignment of the two orbits (see Fig.~\ref{img:sketch}). 
The argument of pericentre and the orbital eccentricity of both coorbital planets are identical, where the first planet starts 
at the pericentre of the common orbit and the second one at the apocenter.
Most of the time this results in
planets on a horseshoe orbit configuration. Other simulations showed that different values for the initial true anomalies of the planets may result in a different kind of coorbital configuration (Tadpole orbit). However, studying our fiducial system revealed the horseshoe configuration to be the most interesting, therefore we
focused the study on that (see also Sec.\ref{sec:discuss}).

Since the potential of the binary differs from a single object, the initial velocities and semi-major axes of the planets are constructed considering the centre of mass rather than the exact potential and, therefore, can differ from the dynamical ideal solution. However, the difference depends on $\Delta \omega$ and is minor. We discuss the influence in the two planets case in Section~\ref{sec:example}. 
To have a conservative estimate, we generally choose the least stable initial angle of $\pi/2$ according to our parameter studies for planet eccentricities of 0.1, 0.3, and 0.5, which is also in agreement with \cite{2018Quarles}.

\begin{figure}
    \centering
    \resizebox{\hsize}{!}{\includegraphics{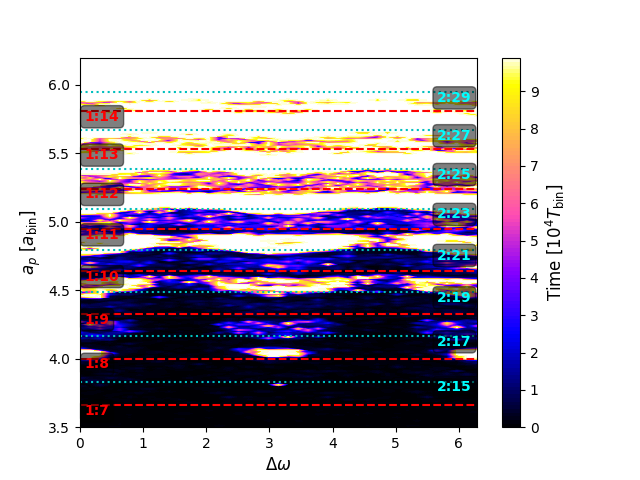}}
    \caption{Stability of coorbital planets with an orbital eccentricity of 0.3 depending on the initial angle between binary periapsis and planet-orbit periapsis. White color shows stable simulations and the colors indicate the survival time of the coorbital configuration. 
    The dashed and dotted lines mark the 1:x and 2:x planet-binary resonances respectively.
    \label{img:periapsis}
    }
\end{figure}

As an example, the $e_\mathrm{p}=0.3$ coorbital case is displayed in Fig.~\ref{img:periapsis}. The effect is more significant in the coorbital configuration.

The planet masses are equal and have a combined mass of $10^{-4}\,M_\mathrm{bin}$, or ${\sim}30\ M_\oplus$. This mass is typical considering the planets observed in a binary system from the Kepler mission.
To resolve the stable orbits we simulated all orbits with a spacing of $\Delta r_\mathrm{peri} = 0.01\, a_\mathrm{bin}$ starting at a clearly unstable orbit $\geq2\,a_\mathrm{bin}$.

\subsection*{Simulation setup}

To simulate each binary system, we use the \texttt{rebound} n-body code by \cite{2012Rein} with the IAS15-integrator \citep{2015Rein} for a 2D setup\footnote{An example setup can be found in \url{github.com/Stefan-Adelbert/1to1circumbinary}.}.

To evaluate the stability, we define the following termination conditions:

The planetary semi-major axis changes substantially. For a single planet, the maximum allowed variation is 20\% of its initial value. For 1:1 planets, the maximum allowed difference between planetary semi-major axes is 20\% of the inner planet's semi-major axis evaluated at each time step.

Coorbital configurations must maintain a nearly identical semi-major axis as a condition for the 1:1 resonance.
As objects on a horseshoe orbit can get closer than $0.1\,a_\mathrm{bin}$ despite the orbit staying stable, we do not use a close encounter termination condition. 
However, as all objects are considered point masses, any collision would lead to an ejection, violating the above condition.
We consider a simulation stable if it reaches $10^5~T_\mathrm{bin}$ without termination. 

\section{Stability of single and 1:1 resonant planets around Kepler 413}\label{sec:example}

\begin{table}[t]
\centering
\begin{tabular}{|c|c|c|c|c|}
\hline
   & $x$ & $y$ & $v_x$  & $v_y$ \\
\hline
Star 1 &-0.384&$1.64\cdot 10^{-4}$&$1.35\cdot 10^{-5}$&-0.4163\\
\hline
Star 2 &0.5760&$1.64\cdot 10^{-4}$&$1.35\cdot 10^{-5}$&0.6245\\
\hline
Planet 1&$3\cdot 10^{-16}$&3.815&-0.5837&$2\cdot 10^{-16}$\\
\hline
Planet 2&-$1\cdot 10^{-15}$&-7.085& 0.3143& 0.0\\
\hline
\end{tabular}
\caption{The exact Cartesian initial values for all objects in the $a_\mathrm{p,ini}=5.45$ setup with $e_p=0.3$ and $\Delta\omega=\pi/2$. 
}
\label{tab:coord}
\end{table}

First, we highlight the stability of single and 1:1 resonant planet systems around a binary like Kepler 413, 
because such a low eccentric binary causes an eccentric disc \citep{2019Kley, 2020Ragusa, 2020Munoz, 2021Penzlin} that more easily traps planets into the 1:1 resonance \citep{2019Penzlin}. With this example, we explore some of the caveats and the properties to consider for the dynamics of the planet pair to choose a setup that does not overestimate the stability.

For the single planet configuration, the planet is added in a Keplerian orbit around the center of mass of the binary.
To initialize the complete 1:1 resonance configuration, we add the second planet in the same Keplerian orbit calculated around the center of mass of the entire single planet configuration, but with a true anomaly of $\pi$, meaning in its apocenter.
The resulting initial coordinates for the example of $a_\mathrm{p,ini}=5.45$ are given in Table~\ref{tab:coord}.

\begin{figure}
    \centering
    \resizebox{\hsize}{!}{\includegraphics{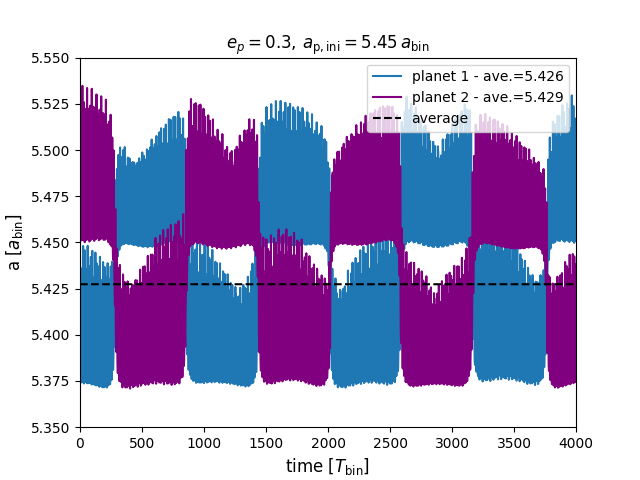}}
    \caption{Semi-major axis evolution of the two planets for a time window of 4000~binary orbits after initialization, showing the libration of the planets around a common orbit. The initial semi-major axis was $5.45\,a_\mathrm{bin}$. However, the average semi-major axis (black dashed line) during the simulation denoted in the top right corner is slightly smaller.
    }
    \label{img:semi}
\end{figure}

Due to the sequential initialization of the planets, the mean orbit can vary in semi-major axis from the initially set value axis by up to $2\%$ as shown in Fig.~\ref{img:semi}.
The alignment becomes a factor due to the non-perfect initial configuration of the planets that also does not include the higher order structure of the binary potential. As discussed before, the initial relative argument of periastron between eccentric planet orbit and eccentric binary orbit affects stability as shown in the example case in Fig.~\ref{img:periapsis} for $e_p=0.3$. 
Compared to a single planet, the coorbital configuration is much less stable and more likely to be interrupted by binary-planet resonances, since the libration of the 1:1 resonant planets leads to natural variations in the semi-major axis of both planets, as described in \cite{2006Mudryk}.
In Fig.~\ref{img:periapsis} the lines mark the positions of the 1:x and 2:x resonances with the binary.
The inner edge of some stability regions (light colors in Fig. 2) coincides with 1:x resonances, while the 2:x resonances limit the outer edges.
The width of these resonant instability regions decreases for increasing semi-major axis as the resonances also become narrow, weakened by the overlap of more higher order resonances and the reduced gravitational potential of the fast-moving stars on the planets. Thus orbits further out become stable.

In general, the stability variations due to the misalignment between binary and planet orbits $\Delta \omega$ is naturally decreased by reducing the eccentricity of the planet's orbit.
Varying the initial orbital distance leads to varying orbital stability. For less eccentric planet orbits, the closest distance between planet and binary varies less with $\Delta \omega$ and, in case of a circular planet orbit, the closest distance remains constant at the difference between planet orbit radius and binary pericenter.
We confirm the findings by \cite{2018Quarles} that initially aligned configurations 
allow more systems to settle into apparent stable states
than those with perpendicular oriented arguments of periastron. So for $\Delta \omega = \pi/2$ the unstable orbits are maximal, therefore we use $\Delta \omega = \pi/2$ in the eccentricity runs to have a conservative measure of the stability of the orbits in the parameter study.  

\begin{figure}
    \centering
    \resizebox{\hsize}{!}{\includegraphics{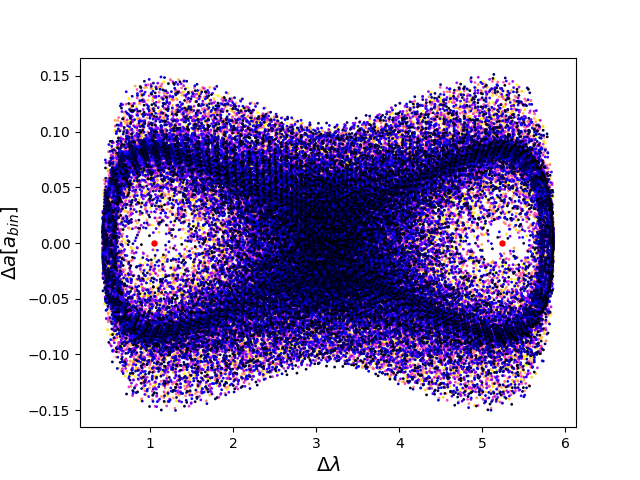}}
    \caption{Plot of the relative mean longitude and semi-major axis between the coorbital planets at $a_\mathrm{p}=5.7\,a_\mathrm{bin}$ with $e_\mathrm{p}=0.3$, indicating that the planets' relative motion describes a stable horseshoe orbit around the L4 and L5 points marked with red circles, without orbit crossing. The color progresses over $10^5~T_\mathrm{bin}$ from yellow to dark blue.
    \label{img:horseshoe}
    }
\end{figure}

\begin{figure}
    \centering
    \resizebox{\hsize}{!}{\includegraphics{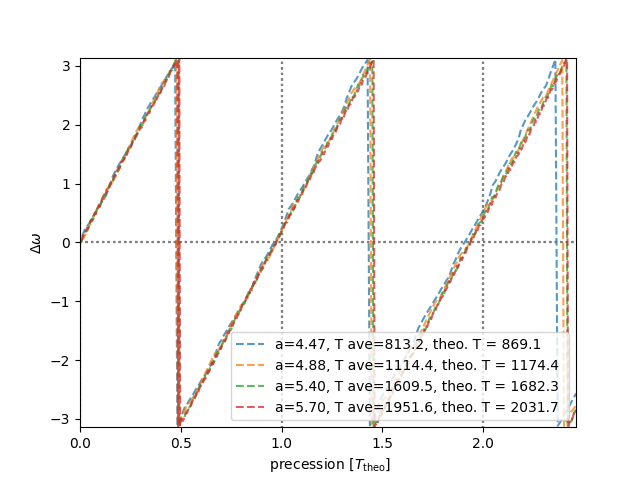}}
    \caption{The relative argument of periastron of the common planets orbit and the binary over the precession rate normalized to the theoretical precession rate from the restricted three-body problem.
    }
    \label{img:prec}
\end{figure}

The relative orbits between the planets are generally horseshoe orbits as shown in Fig.~\ref{img:horseshoe} for the example of $e_p=0.3$, $a_p=5.7\,a_\mathrm{bin}$. The planets librate around the Lagrange point locations L4 and L5 and exchange their angular momentum with the orbit swap (see Fig.~\ref{img:semi}). The pattern is more complex due to the effects of the binary, however, it is periodic, as the color scale from yellow to blue tracks the linear time progression of $10^5~T_\mathrm{bin}$.
To investigate the dynamics of the planets we compare our findings with the expected precession period from the restricted three-body problem.
Figure~\ref{img:prec} shows the precession of the common planet orbit for a planet eccentricity of $e_p=0.3$. The time is normalized to the theoretical precession of the restricted three-body problem derived from \cite{2004Moriwaki}
\begin{equation}
T_\mathrm{prec} = \frac{4}{3} \frac{(q+1)^2}{q} a^{3.5} \frac{(1-e_p^2)^2}{1+1.5 e_\mathrm{bin}^2} T_\mathrm{bin}. 
\end{equation}
In the selected cases, the precession of the resonant planets is systematically slightly slower by $\approx 50-70\,T_\mathrm{bin}$ for stable orbits. However, this precession period is still close to the prediction of the restricted three-body problem.
This does not change by reducing the planet masses to $10^{-6}\,M_\mathrm{bin}$.

\begin{figure}
    \centering
    \resizebox{\hsize}{!}{\includegraphics{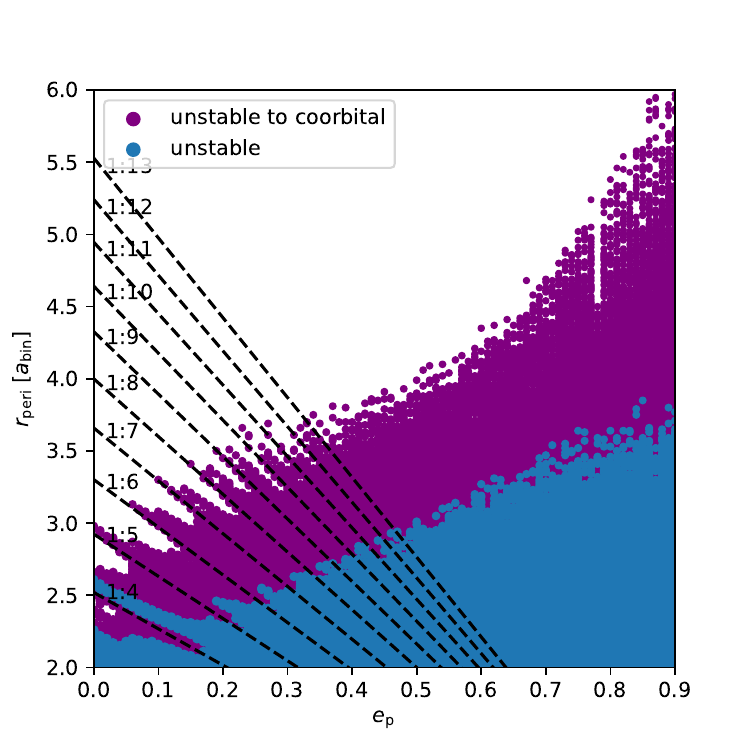}}
    \caption{Map of the survival of single and double (coorbital) planet systems with varying planet eccentricities $e_p$. Blue is unstable for all, purple is only stable for single planets but not the coorbital configuration. The black lines mark 1:x resonances between x=4--13 from shallowest to steepest.
    \label{img:ecc_both}
    }
\end{figure}
\begin{figure}
    \centering
    \resizebox{\hsize}{!}{\includegraphics{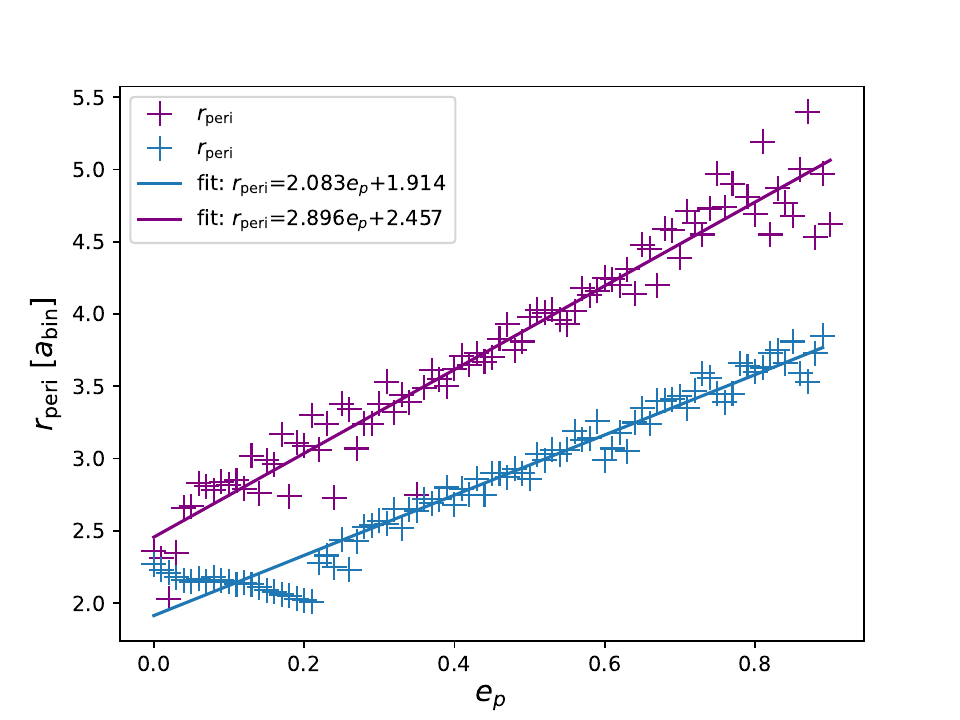}}
    \caption{A linear fit of the pericentre of the first stable orbits in Fig.~\ref{img:ecc_both}. The values close to $e_\mathrm{p}=0$ diverge from the plot due to a unstable region around the 1:4 resonance. 
    \label{img:Fit1}
    }
\end{figure}

To compare to previous work we simulate a range of single circumbinary planet systems and vary the initial planetary eccentricity and semi-major axis.  
Fig.~\ref{img:ecc_both} shows the 
instability transition with increasing planetary eccentricity and correspondingly increasing pericentre distance.
The non-blue region denotes stable initial conditions for a single planet (up to $10^5\ T_{\rm bin}$).
Fig.~\ref{img:ecc_both} demonstrates that with increasing eccentricity the first stable orbit moves further out for the single planet. By plotting the pericentre distance of the first stable planet orbit a linear trend can be found, with the exception of very small planet eccentricities. Due to the 1:4 resonance, the values for $e_\mathrm{p}<0.2$ do not follow the overall upward trend of instability. The 1:4 resonance corresponds to the lowest stable region in Fig.~\ref{img:ecc_both} surrounding the lowest black dashed line plotted. 
The effect of the 1:x resonances in the single planet case beyond 1:8 becomes hardly noticeable.
The critical semi-major axis for the circular planet of $2.3\,a_\mathrm{bin}$ compares well with the value for a circular planet around a circular binary in \cite{1999Holman}. Fig.~\ref{img:Fit1} shows the fitted trend of stability for the single planet in blue. Due to resonance some orbits further out are unstable (see Section~\ref{sec:resonance}), however, for the more general trend using the first stable orbit is sufficient. The fitted stability limit is
\begin{equation}
 r_\mathrm{c,peri} = [(2.08 \pm  0.10)\, e_p + (1.91 \pm 0.05)]\, a_\mathrm{bin}, 
\end{equation}  
using standard deviation errors.

The simple linear fit describes the stability behaviour of a single planet around a Kepler-413 like binary fairly well for $e_\mathrm{p}>0.2$. However, the fit ignores the resonant unstable lines. Especially the 1:4 and 1:3 resonant instability decrease the stability for smaller planet eccentricities ($e_\mathrm{bin}<0.1$) below a linear fit. The fit also slightly overestimates the stability for the most eccentric systems, which is realized in other works making use of a quadratic term \citep[e.g.][]{2001Mardling}.

Now focusing on the results of the coorbital planet configuration, Fig.~\ref{img:ecc_both} shows the stability of orbits for different planet eccentricities and pericentre positions up to an eccentricity of 0.9 in purple. Stable orbits exist for all eccentricities if the orbit is sufficiently large. 

For the 1:1 resonant planets as well, the stable orbits follow a linear trend, plotted in Fig.~\ref{img:Fit1}. The linear fitting function is given by: 
\begin{equation}
r_\mathrm{c,peri}= [(2.90 \pm 0.15) e_\mathrm{p} + (2.46 \pm 0.08)]\, a_\mathrm{bin}.
\end{equation} 
The offset of this function here is larger than for the single planet, meaning that stable orbits in the case of 1:1 resonant planets around a binary start $>0.5\,a_\mathrm{bin}$ further out, and the function for 1:1 resonant planets is steeper with planet eccentricity. 
So single planets with eccentric orbits can survive much closer to the binary than 1:1 resonant planet configurations.
This is a result of the broadened libration of the planet orbits due to the added coorbital angular momentum exchange.

However, stable 1:1 resonant planets around binaries are possible within a pericentre of $r_\mathrm{peri}< 5.6\,a_\mathrm{bin}$. This means that the planets still survive on orbits that pass by close to the binary even for extremely eccentric 1:1 configurations.

\section{Range of stable 1:1 resonance\label{sec:resonance}}

\begin{figure*}
    \centering
    \resizebox{\hsize}{!}{\includegraphics{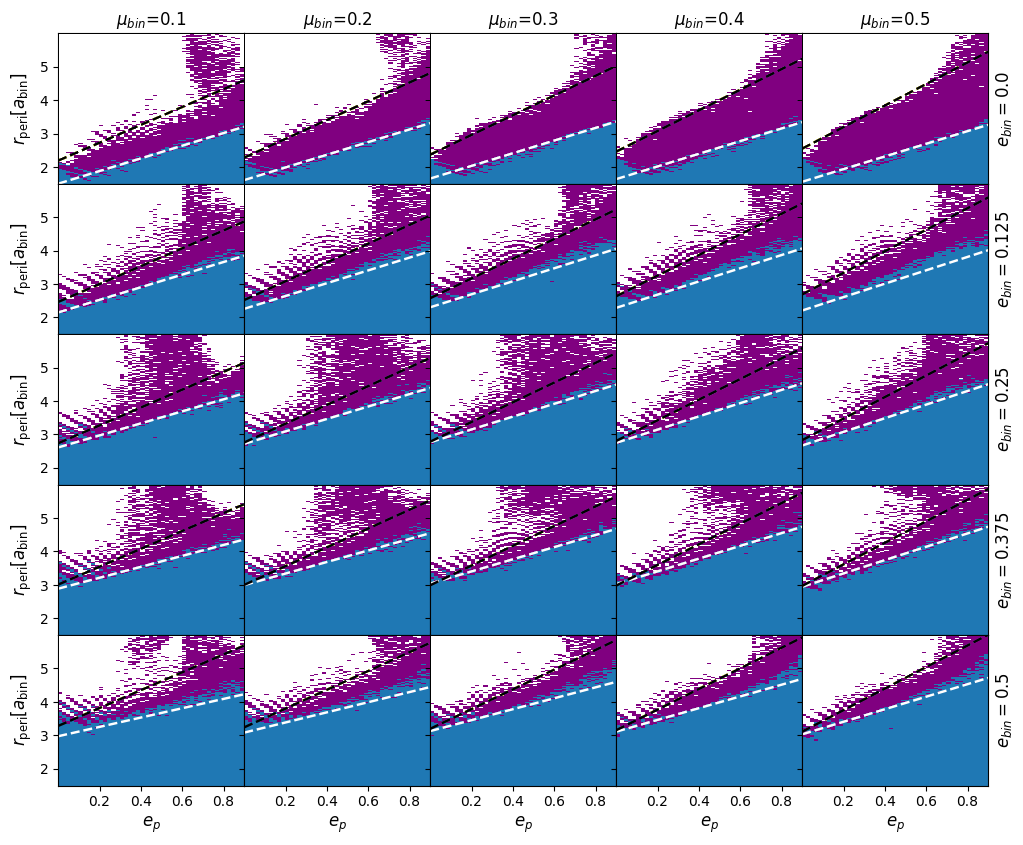}}
    \caption{Stability for a parameter range in binary eccentricity (rows) and binary mass ratio (columns) for planet eccentricities against orbit pericentre. Blue indicates unstable orbits, while magenta indicates unstable coorbital conditions. The lines represent the best fit to the first stable orbits.
    \label{img:range2}
    }
\end{figure*}

After using this specific example of a Kepler 413-like system, we vary the binary parameters to compare the results to the general stability of planets around binaries as investigated before for the single planet case \citep{1986Dvorak,1999Holman,2015Chavez,2016Popova,2018Quarles} and create a simple model of the stability limits around binary stars.
For this, we change the mass fraction $\mu=\lbrace 0.1,0.2,0.3,0.4,0.5\rbrace$, while also changing the binary eccentricity $e_\mathrm{bin}=\lbrace 0,0.125,0.25,0.327,0.5\rbrace$. We limited the range of binary eccentricities in this study to the range of the observed host binary stars, with Kepler 34 being the most eccentric, planet-hosting binary with $e_\mathrm{bin}=0.52$ \citep{2012Kepler34-35}. 
In Fig.~\ref{img:range2} we show the results of the orbital stability for the whole data set. The color marks the logarithmic survival time of the system, where white systems survived $10^5~T_\mathrm{bin}$. More equal mass binaries reduce the survival of more eccentric planet configurations. More eccentric binary stars shift the first stable orbit in general further out as is also  suggested in other work on single planets \citep[e.g.][]{2015Chavez,2016Popova,2018Quarles}. Interestingly, the moderately binary eccentricities lead to unstable regions further out after the first stable orbit for configurations with planets more eccentric than 0.5. For eccentric binary systems, the resonant orbits cause even more pronounced unstable feature lines. 

Comparing the single planet and coorbital planet configurations, the slope of the unstable region is steeper for coorbital configurations, especially with increasing binary eccentricity as also discussed in Sec.~\ref{sec:example}. Meanwhile, the first stable orbit between the single and the coorbital planet on circular orbits become much more similar with increasing binary eccentricity.
In general, we can create a good first order fitting relation by considering a linear function of the first stable pericentre against planet eccentricity as already used in Sec.~\ref{sec:example} for the coorbital configuration. The strength of the resonant features will worsen such a fit, especially, when both eccentricities are low for the single planet case.

We based our fit on the quadratic fit function by \citep{1999Holman} and extended it by the planet eccentricity. We reach a best fit using a non-linear least square to the function. For this we take the first stable orbit as a function set and the combination of [$e_\mathrm{bin}, \mu_\mathrm{bin}$, $e_\mathrm{p}$] as data set to solve for the coefficients assuming a 2nd order term approximation can describe the limits of stability. The mixed terms, $e_\mathrm{bin}\mu_\mathrm{bin}$, $e_\mathrm{p}\mu_\mathrm{bin}$ and $e_\mathrm{p}\mu_\mathrm{bin}^2$ are dropped as they increase complexity without model improvement.
Thereby we reach a best fit for the single planet case in the following form:
\begin{align}
r_\mathrm{c,peri}&/ a_\mathrm{bin} = (1.36 \pm 0.05) \\ \nonumber
+& (5.79 \pm 0.19) e_\mathrm{bin} - (5.87 \pm 0.39) e_\mathrm{bin}^2 \\ \nonumber
+& (1.99 \pm 0.32) \mu_\mathrm{bin} - (3.14 \pm 0.52) \mu_\mathrm{bin}^2\\ \nonumber
+& [(1.85 \pm 0.05) \\ \nonumber&- (2.10 \pm 0.46) e_\mathrm{bin}^2 + (3.00 \pm 0.52) e_\mathrm{bin} \mu_\mathrm{bin}] e_\mathrm{p}
\end{align}

The overall fit reaches a coefficient of determination of $R^2=0.962$, which is a fair correlation for such a limited data set with 3 independent variables.
However, due to the reduction of the terms the remaining coefficients deviated by factor of at most 4.5 from the values found by \cite{2018Quarles}, while describing comparable data.
 When comparing our zero eccentricity planet case to the study of circular planet orbits by \citep{1999Holman} this leads to a significant deviation, as it ignores the strong resonances acting in circular binaries on circular planets that bend the stability region outwards for planet eccentricity $e_\mathrm{p}<0.2$ as noticeable in Fig.~\ref{img:range2} in the data and a deviation seen in the example fit of Fig. \ref{img:Fit1}.
 The same resonant lines of instability reaching further pericentre distances have been also observed by e.g. \cite{2015Chavez, 2016Popova} and \cite{2018Quarles}.

\begin{figure*}
    \centering
    \resizebox{\hsize}{!}{\includegraphics{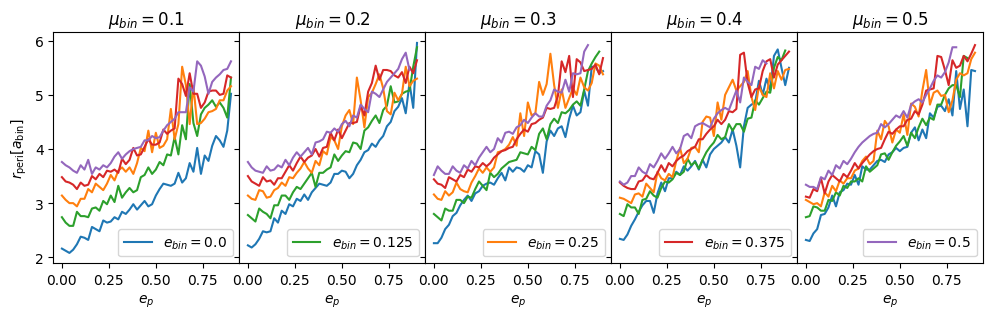}}
    \caption{First stable coorbital configurations depending on the planet eccentricity. From left to right the mass ratio $\mu$ increases towards equal mass. The color indicates varying binary eccentricities from circular (blue) to $e_\mathrm{bin}=0.5$ (purple).
    \label{img:range_crit}
    }
\end{figure*}

In Fig.~\ref{img:range_crit}, the first stable coorbital configuration is shown to move further out with increasing binary eccentricity with linear spacing. 
Changing the binary masses to equal values reduces the difference caused by the binary eccentricity and steepens the slope towards more distant stable orbits for more eccentric planet orbits.
However, the stronger non-linear instabilities for $e_\mathrm{p}>0.5$ make the resulting curve noisy in this region.
Nevertheless, from the data we fit a very simple estimate on the stability limits given $\mu,e_\mathrm{bin}$ and $e_\mathrm{p}$. As our sample size in  $\mu$ and $e_\mathrm{bin}$ is small we used a similarly simple form of equation as \citep{1999Holman} to fit these values to avoid overfitting. 
We again reduced the equation by all terms that did not produce a contribution above the level of their error after the first fitting iteration. After doing this we arrive at the following simplistic equation to predict an estimate for the first stable orbit for the two-planet coorbital configuration:

\begin{align}
 r_\mathrm{c,peri}&/ a_\mathrm{bin} = (2.10 \pm 0.08) + (2.54 \pm 0.12) e_\mathrm{p} \\ \nonumber
 &+ (2.45 \pm 0.18) e_\mathrm{bin} + (0.90 \pm 0.23) \mu_\mathrm{bin} \\ \nonumber
&+ ( 1.37 \pm 0.36) e_\mathrm{p} \mu_\mathrm{bin} + (- 2.65 \pm 0.54) e_\mathrm{bin} \mu_\mathrm{bin} \label{eq:fullfit}
\end{align}

\begin{figure}
    \centering
    \resizebox{\hsize}{!}{\includegraphics{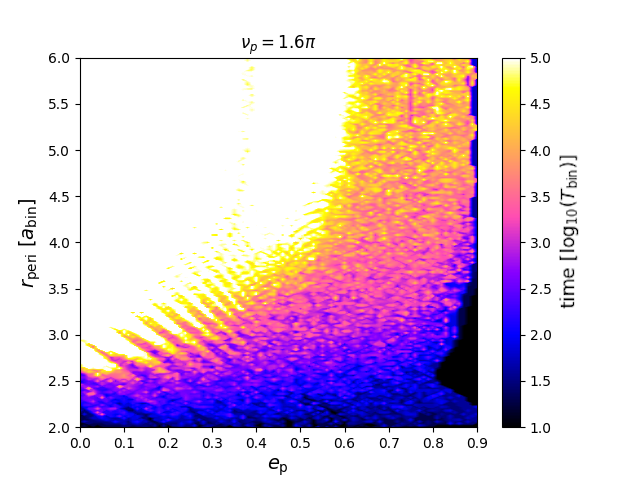}}
    \caption{Stability of Tadpole configuration for the Kepler-413 example case. The color scale indicates the survival time of the planet configuration.
    \label{img:tad}
    }
\end{figure}

The fit reaches a coefficient of determination of $R^2=0.932$ for the limited parameter set in the binary parameters. As the first orbit is nearly linear, this is a good approximation for the realm of stability of  eccentric, circumbinary, coorbital configurations.
However, as this is just a simple fit, it does not take into account the unstable lines created by the resonances between binary and planet or the unstable regions beyond the first stable orbit for higher planet eccentricities and is only relevant for horseshoe orbit configuration of planets. We investigated some examples of tadpole orbits by initializing the planets near the L4 and L5 points. In the case of tadpole orbits the coorbital planet configuration for all tested parameters grow unstable for $e_\mathrm{p}>0.6$ as shown in Fig.~\ref{img:tad}. At this point, all planet configurations with orbit eccentricities $e_\mathrm{p}>0.5$ should be looked at with caution, while a stability limit for lower eccentricities can be predicted based on our simulations. 

\section{Discussion}\label{sec:discuss}

Finding horseshoe orbits to be stable is well in agreement with the recent theoretical findings of coorbital multi-planet systems around single stars by \cite{2023Raymond}. However, especially the regions of resonance with the binary cause instabilities that extend further for coorbital planet systems than for single planets as the momentum exchange between the planets causes them to enter into the unstable orbit while completing the horseshoe revolution.

The disruption effects of this exchange are comparable to the tidal interaction studied by \cite{2001Mardling}. Coorbital resonance is slightly more effective in destabilizing by broadening the effective orbit to the common horseshoe orbit (see e.g. Fig.~\ref{img:semi}\&\ref{img:horseshoe}). This shifts the first stable orbit in the coorbital case compared to the tidal interaction case out by about $\sim0.1~a_\mathrm{bin}$ for the circular binary. 
The simulations are limited to two dimensions only. However, most observed, close circumbinary planets are well aligned with the plane of the binary, thereby, our case is the most common picture.
The transit observations are biased towards finding primarily well-aligned planets.
However, \cite{2020Chen,2023Chen} showed that the polar planet configuration is more stable than the aligned configuration, therefore our results are conservative cases.

In this study, only the stability of planets that are already initialized in a coorbital configuration is tested without the influence of hydrodynamics. This represents only the final planetary system and does not explain the formation or migration of planets leading to such configurations.
The smooth transition from planets embedded in a disc, getting trapped into resonance near the inner cavity, and the disc eventually dissipating would be one of the next steps. 
In future studies, it would allow us to understand if the 1:1 configuration is affected by the gas dissipation. But the longer simulation time for the hydrodynamic disc model and the n-body is beyond the scope of this work.
\cite{2022Martin} investigated the destabilization of single planets during the migration using a simplified static disc model. Their results show paths for stable migration of planets of the mass used in this study to the close-orbits up to the 1:6 resonance with the binary.
In \cite{2019Penzlin}, we found 
using viscous 
hydrodynamical disc simulations that the inner density maximum close to the inner cavity can provide a high pressure environment which damps migration planets into a 1:1 resonance for nearly circular binaries. 
\cite{2023Coleman} improved the disc description to include eccentricity used in their n-body investigation and found, in fact, some coorbital planets forming. 
Here, we can show even if the orbits get closer while the disc gets colder and less dense there are still coorbital configurations possible.

However, especially the stable orbits closest in are sensitive to the initial conditions used. Changing the initial conditions like the precise angle between the pericentre of planets and stars can affect the stability for all simulations of systems with three or more bodies, especially since precession is slow.

For a conservative estimate, the least stable angle was chosen \citep[see also][]{2018Quarles}.
While for low eccentricity the first stable orbits allow still for more than 20 full precessions during the simulation time, for $e_p=0.9$ the semi-major axis of the first stable orbit is so wide that the planets can only complete about one full precession during the simulation. Thereby the initial configuration can become important.
While for a single planet case the long-term stability, especially for these very eccentric planet orbits, could be better determined by a chaos indicator \citep[see e.g.][]{1997FLI,2000RLI}, this is not restrictive enough to distinguish between coorbital and non-coorbital two planet systems.

\cite{2015Chavez} and \cite{2016Popova} investigate the stability for a number of the observed circumbinary planet systems. 
For low eccentricity binaries (e.g., Kepler-47 in the \cite{2016Popova} study) or the Kepler-413 case \citep{2015Chavez,2018Quarles}, the distribution of stable orbits becomes nearly linear especially for $e_\mathrm{p}<0.8$ and matches well with our finding. Comparing to the classic work of \citep{1999Holman}, as we included 3 variable parameters we simplified the fitting function to a lower order. \cite{1999Holman} will be more accurate for circular planet configurations. 

When attempting to fit in the same form for the single planet case, we found some deviations with our model, that is strongly informed by eccentric planets.
Our fit is generally below their solution by $\sim 0.5~a_\mathrm{bin}$ which also is evident from the top row of Fig.~\ref{img:range2}. The fit ignores the effects of the resonance. Thus with our $e_\mathrm{p}$-dependent fit, we reach a stability limit that is further in, especially for low planet eccentricities that are disturbed by the most by the resonances. \cite{2018Quarles} gives a good measure for the stability of eccentric single planets taking the strong resonant features into account.
Given the focus on eccentric planet orbits in this work, the fit is aimed at a regime where the resonant instabilities lose significance, which is also a natural result for the coorbital planets as they start to become stable further out where the effects of resonances dominate the stability map less.

Previous studies \citep{2018Quarles, 1999Holman, 2016Popova, 2019Chen} on single planet stability around binaries have considered a range of binary eccentricities.
However, for the 1:1 resonant planet case we put emphasis on nearly circular binaries here, since in \cite{2019Penzlin} we found a hydrodynamical mechanism to form 1:1 resonant planets by migration only around nearly circular binaries. Further, we only consider a limited parameter range to create some general idea about the binary dependent behaviour while rather focusing on the planet eccentricity as a parameter.
\cite{2022Fitzmaurice} simulated the migration of multiple planets using \texttt{rebound} with a prescribed disc. Their models achieve coorbital configuration rarely with the circular disc torque used. Such a circular disc torque would in full hydrodynamics simulations occur for more eccentric binaries \citep{2021Penzlin} than the case studied. However, coorbital resonances are not impossible also for more eccentric host binaries.

\section{Conclusion}\label{sec:results}
In this work, we investigate the stability of coorbital planets compared to single planets around binary stars. In the example of a Kepler-413 with one planet, we confirm the previous analysis of circumbinary planet stability \citep{2018Quarles}. In addition, we show that stable coorbital, circumbinary planet configurations are possible even for eccentric planet orbits.

As a good approximation of the 1:1 stability around a Kepler 413 analogue, we found the simple relation of $r_\mathrm{c,peri}= [(2.90 \pm 0.15) e_\mathrm{p} + (2.46 \pm 0.08)]\, a_\mathrm{bin}$. 
The stable zone for 1:1 resonant planets is thereby further out than in the single planet case. The critical pericentre for the coorbital configuration is $\geq 0.8~a_\mathrm{bin}$ farther out than in the case of a single planet, with even greater differences for more eccentric orbits. However, the stability limit of coorbital configuration is still close to the binary and the orbits achieved by migration in \cite{2019Penzlin} are stable for $>10^5\, T_\mathrm{bin}$.

When investigating a range of binary mass ratios and eccentricities, we find that the linear $r_\mathrm{c,peri}= (A\, e_p + B)\, a_\mathrm{bin}$ fit remains a reasonable fit to the data. And with this we can confine stability limits that depend on all 3 free parameters, $\mu_\mathrm{bin}$, $e_\mathrm{bin}$ and $e_\mathrm{p}$ with a very simple first order fit.
It is noteworthy that the retrieved stability is specifically valid for coorbital planets in a horseshoe configuration, while tadpole configurations become unstable for planet eccentricity $e_\mathrm{p}>0.5$. For the horseshoe configuration for $e_\mathrm{p}>0.5$ unstable regions beyond the first stable orbit appear. For a planet eccentricity $e_\mathrm{p}<0.5$ and binary eccentricities $e_\mathrm{bin}<0.5$ we find that close-in, stable, coorbital, planet orbits are possible.

\begin{acknowledgements}
In Memory of Willy Kley, who initiated this project. His patient help and kind advice made this possible and we are grateful for every day we worked with him.
The authors acknowledge support by the High Performance and Cloud Computing Group at the Zentrum für Datenverarbeitung of the University of Tübingen, the state of Baden-Württemberg through bwHPC
and the German Research Foundation (DFG) through grant no INST 37/935-1 FUGG. 
This research was supported by the Munich Institute for Astro-, Particle and BioPhysics (MIAPbP) which is funded by the Deutsche Forschungsgemeinschaft (DFG, German Research Foundation) under Germany´s Excellence Strategy – EXC-2094 – 390783311.
All plots in this paper were made with the Python library matplotlib \citep{Hunter:2007}.
AP acknowledgs support from the Royal Society 2021 Research Fellows Enhanced Research Expenses RF/ERE/210064 and Enhanced Expenses Award and grant KL 650/26 of the German Research Foundation (DFG).
\end{acknowledgements}

\bibliography{1to1nbody}
\bibliographystyle{aa}

\end{document}